\documentclass[a4paper,preprintnumbers,floatfix,superscriptaddress,aps,pra,twocolumn,showpacs,notitlepage,longbibliography]{revtex4-1}
\usepackage[utf8]{inputenc}
\usepackage[T1]{fontenc}
\usepackage[sc,osf]{mathpazo}
\usepackage{amsmath, amsthm, amssymb,amsfonts,mathbbol,amstext}
\usepackage{graphicx}
\usepackage{dcolumn}
\usepackage{bm}
\usepackage{bbm}
\usepackage{hyperref}
\usepackage{mathtools}
\usepackage{comment}
\usepackage{color}
\usepackage{dsfont}
\usepackage[normalem]{ulem}
\usepackage{multirow}
\usepackage{diagbox}
\usepackage{pdfcomment}

\definecolor{applegreen}{rgb}{0.55, 0.71, 0.0}
\definecolor{darkelectricblue}{rgb}{0.03, 0.51, 0.57}
\definecolor{forestgreen}{rgb}{0.13, 0.55, 0.13}
\definecolor{atomictangerine}{rgb}{1.0, 0.6, 0.4}

\newcommand{\phs}{\phantom{+}}

\def\1{\mathbf{1}}
\def\0{\mathbf{0}}

\def\Id{\mathbbm{1}}

\DeclareMathOperator{\tr}{tr}
\DeclareMathOperator{\Tr}{Tr}

\begin{document}
\title{Multistage games and Bell scenarios with communication}
\author{George Moreno}
\affiliation{International Institute of Physics, Federal University of Rio Grande do Norte, 59070-405 Natal, Brazil}
\author{Ranieri Nery}
\affiliation{International Institute of Physics, Federal University of Rio Grande do Norte, 59070-405 Natal, Brazil}
\author{Alberto Palhares}
\affiliation{Departamento de Física Teórica e Experimental, Universidade Federal do Rio Grande do Norte, Natal, RN, 59078-970, Brazil}
\author{Rafael Chaves}
\affiliation{International Institute of Physics, Federal University of Rio Grande do Norte, 59070-405 Natal, Brazil}
\affiliation{School of Science and Technology, Federal University of Rio Grande do Norte, 59078-970 Natal, Brazil}

\date{\today}
\begin{abstract}
Bell nonlocality is a cornerstone of quantum theory with applications in information processing ranging from cryptography to distributed computing and game theory. Indeed, it is known that Bell's theorem can be formally linked to Bayesian games, allowing the use of nonlocal correlations to advise players and thereby achieve new points of equilibrium that are unavailable classically. Here we generalize this link, proving the connection between multistage games of incomplete information with Bell scenarios involving the communication of measurement outcomes between the parties. We apply the general framework for a few cases of interest and analyze the equilibrium reached by quantum nonlocal correlations.
\end{abstract}

\maketitle
\section{Introduction}
Game theory \cite{fudenberg1991} provides a systematic way of evaluating the unrolling of events involving a number of rational individuals (players) actively participating on it and having their own interests on the possible outcomes. Given some available information and a range of situations they can foresee, rational players will never take decisions leading to a less desirable outcome. Such analysis has found applications in the most diverse fields of knowledge \cite{aumann1992handbook,han2012game,colman2013game}. In the late 90's quantum versions of it, that is, games in which players share quantum states and actions are translated into unitary operations, were first presented \cite{Meyer1999,Eisert1999}. An active and diverse literature has then been established, connecting quantum correlations to various forms of games \cite{Benjamin2001,Eisert2001,Chen2002,Du2002,Iqbal2002,Flitney2002,Du2003,d2002quantum,piotrowski2003invitation,Iqbal2008,Flitney2009,Iqbal2010,Pappa2015,Roy2016,rai2017strong,banik2019two,Brunner2013}. 

More recently, based on the similarity between the game's rules and the concept of local realism in Bell's theorem \cite{bell1964einstein}, Brunner and Linden established a connection between correlated Bayesian games and Bell nonlocality \cite{Brunner2013}. As a result, classical advisors are limited by the constraints of local hidden variable models, whereas correlations violating a Bell inequality may lead to new possible payoffs and new Nash equilibria as well. That is, strategies based on quantum correlations lead to scenarios in which no rational player would individually change their actions. Variations of this game featuring fair and unfair strategies, as well as conflicting interests, were also considered \cite{Pappa2015,Roy2016}. These results encompassed earlier works based on nonfactorizable joint probabilities \cite{Iqbal2008} and other approaches involving the concept of Bell nonlocality \cite{Flitney2009}. At the same time, following Ref. \cite{Iqbal2010}, such works circumvented criticisms on the early formulation of quantum games \cite{Benjamin2001,Eisert2001}, since the quantum advantage is achieved in a device-independent setting and irrespectively of any details or constraints over the quantum states, their evolution or the measurement setup.

In spite of the foundational importance of Bell's theorem \cite{bell1964einstein} and its various applications in quantum information processing \cite{brunner2014bell}, the constraint that the involved players are space-like separated, thus they cannot communicate, turns out to be rather restrictive, especially when dealing with dynamic games, where there is an underlying temporal order between the different players. To circumvent such limitations one has to consider Bell scenarios where some form of communication between the parties is allowed. Originally introduced in the context of witnessing genuine multipartite nonlocality \cite{Svetlichny1987,Collins2002,Jones2005,Bancal2009} and typically referred to as nonlocal hidden variable models, such Bell scenarios with communication have later been used to understand how to simulate quantum correlations with classical models where the locality assumption in Bell's theorem is relaxed \cite{bacon2003bell,pironio2003violations,pawlowski2010non,maxwell2014bell,Chaves2015,ringbauer2016experimental,Brask2017}. More recently, such generalized Bell scenarios have been firmly established \cite{wood2015lesson,Chaves2015,chaves2018quantum,van2019quantum} in connection with some tools and concepts from the field of causal inference \cite{pearl2009causality} also starting to find applications in processing of information \cite{ried2015quantum,agresti2020experimental,Moreno2020,gachechiladze2020quantifying}.

Here, we generalize the connection first made in Ref. \cite{Brunner2013}, establishing a link between multistage games with incomplete information and Bell scenarios with outcome communication. More precisely we show that nonlocal hidden variable models limit the space of payoff functions achievable by an advisor resorting to classical correlations. Bell inequalities can be directly associated with the payoff of the players, in such a way that a quantum violation of these inequalities lead to better game outcomes. Finally, we analyze a few cases of interest involving an arbitrary number of players and show in particular that a quantum advantage is  already possible in the simplest dynamic game, involving only two players.

This manuscript is organized as follows. In Sec. \ref{sec:2}, we make a general introduction to game theory and describe in detail multistage games and its forms of correlated equilibrium. In Sec. \ref{sec:3}, we follow the approach introduced in Ref. \cite{Brunner2013} and establish a formal link between Bell scenarios with outcome communication and multistage games. In Sec. \ref{sec:4}, we explore a few scenarios of interest and show how quantum correlations violating Bell inequalities can lead to better payoffs and new equilibrium in dynamic games. Finally, in Sec. \ref{sec:5}, we discuss our results and point out possible future directions for research.

\section{Multistage games and their form of correlated equilibrium}
\label{sec:2}

In Game theory games are classified according to two criteria: dynamics and information \cite{fudenberg1991}. If the order in which players play does not affect the outcome of the game, that is, the payoff functions, then the game is named \emph{static}, otherwise it is referred to as \emph{dynamic}. On the other hand, if all players have access to all parameters relevant to the execution of the game we call it \emph{game with complete information}, and whenever some parameters are unknown to some player we have a \emph{game with incomplete information}. We thus have four possible kinds of games.  Static games with complete information are known as strategic form games. In turn, a static game with incomplete information is known as a Bayesian game.
A dynamic game with complete information is named a multistage game. Finally, a dynamic game with incomplete information is referred to as multistage game of incomplete information. Multistage games of incomplete information will be the focus through this work.

A multistage game of incomplete information consists of a set of players $N=\{1,...,n\}$, a set of types $\Theta = \bigtimes_{i=1}^n\Theta_i$, a set of actions $A = \bigtimes_{i=1}^n A_i$, and a set of stages $\{1,...,K\}$. Types are sorted according to a distribution $p(\theta)$, which we refer to as the prior distribution, where $\theta=(\theta_1,\dots,\theta_n) \in\Theta$, and such that $p(\theta) = \prod_{i=1}^n p(\theta_i)$ (that is, the types are uncorrelated). Each player has information only about its own type (private information), never on the other players type. Here, we will focus on a sequential game of $K = n$ stages, in which each player $i$ plays alone in stage $k = i$ revealing its action $a_i$ at the end of the stage. Given $k$, the most general history of all actions previous to $k$ is represented by the set of actions $(a_{k-1},...,a_{1})$. Notice, however, that in some multistage games the history might be a subset of all previous actions. For example, in an open auction the only relevant action for a player at stage $k$ is the directly preceding one and in this case the history is simply given by $a_{k-1}$. We will call $h_i$ the history accessible to player $i$ at the stage $k=i$. In the situation we consider here, the most general payoff function for each player is then $u_i(a,\theta)$ where $a=(a_1,\dots,a_n)$.  We highlight that a Bayesian game, as the one considered in Ref. \cite{Brunner2013}, corresponds to a static game, i.e., $K=1$.

A central concept in Game Theory is that of equilibrium. The most iconic form of equilibrium being the Nash equilibrium \cite{Morgenstern1953, Nash1951}, which is also the basic concept behind more sophisticate forms of equilibrium, e.g., Bayesian Nash equilibrium \cite{kajii1997robustness}, perfect equilibrium \cite{moore1988subgame}, and perfect Bayesian equilibrium \cite{fudenberg1991perfect}. Nash equilibrium is defined as the set of actions chosen by the players in such way that none of them can increase its own profit by individually changing its actions, thus providing most likely outcomes for the game. In this manuscript, we resort to the concept of \emph{correlated equilibrium} (CE), proposed by Aumann \cite{Aumann1974}, and \textit{extensive-form correlated equilibrium} (EFCE) introduced in Ref. \cite{Bernhard2008}.

The concept of correlated equilibrium is built up on the strategic form of games and defines an advisor which provides information for the players to set the strategy they should use. These advisors happen to have the exactly same properties of a shared hidden variable in the context of Bell's theorem \cite{Brunner2013}. More formally, correlated equilibrium considers an autonomous communication device (which we will refer to as canonical communication device or CCD) that recommends a strategy profile for player $i$, via a signal $\lambda$ according to a probability distribution $p(\lambda)$ which is defined before the game starts for each type of player \cite{Bernhard2008}. This implies that the behavior of player $i$ should follow
\begin{equation}
\sigma_i(a_{i}|h_{i},\theta_i,\lambda) = \sigma_i(a_{i}|\theta_i,\lambda). 
\end{equation}
that is, the behavior of the players can only depend on their types but not on the history (actions) of the other players. If the profile of behaviours set by the CCD fits the definition of a Nash equilibrium, i.e., no rational player would disobey the advisor, then it represents a CE.

While this concept suits Bayesian games perfectly \cite{Brunner2013}, it does not contemplate the possibilities of a multistage game where the behavior of players have an intrinsic dependence on the history of the game. To assess this scenario, we consider the concept of an extensive-form correlated equilibrium. The EFCE is defined in a scenario with perfect recall in which an autonomous communication device (which we will name extensive communication device or ECD) sends a signal $\lambda\in\Lambda$, sampled according to a distribution $p(\lambda)$, which sets a strategy profile for each type and each available history on each stage. A system is said to be in an EFCE if it represents a Nash equilibrium of the game obtained by the introduction of the advisor, more commonly referred to as the extended game.

\section{Multistage games as a Bell scenario with communication}
\label{sec:3}

In the following, we will show a link between multistage games and Bell scenarios involving the communication of measurement outputs. The most general history $h_{i}$ available to player $i$ consists of all actions $(a_{i-1},\dots,a_1)$. We will also consider a more general case where the ``memory'' $m_i$ of each player might be limited , that is, each player has access to a partial list of previous actions such that $h_{i}=(a_{i-1},\dots,a_{i-m_i}$).  In particular, if $m_i=0$ player $i$ has no access to previous actions ($h_{i}=\emptyset$, which corresponds to the Bayesian game considered in \cite{Brunner2013}) and if $m_i=i-1$ the player has access to all previous actions.

A possible payoff $U_i$ for player $i$ is given by the following equation:
\begin{eqnarray}
U_i=\sum_{a,\theta} p(a,\theta)u_i(a,\theta),
\end{eqnarray}
in which the distributions $p(a,\theta)$ are set by the structure of the game. For instance, the imposition of limited memory plus the sequential feature leads to distributions of the form:
\begin{eqnarray}
\label{eq: prob decomposition}
    p(a,\theta) = p(\theta)\prod_{j=1}^n \sigma_j(a_j|a_{j-1,\dots,j-m_j},\theta_j),
\end{eqnarray}
here we have introduced the short-hand notation $a_{j-1,\dots,j-m_j} = a_{j-1},...,a_{j-m_j}$.

Thus the payoff function above is given by
\begin{multline}
\label{eq: possible payoff 2}
U_i = \sum_{a,\theta}p(\theta)\prod_{j=1}^n \sigma_j(a_j|a_{j-1,\dots,j-m_j},\theta_j)u_i(a,\theta).
\end{multline}

In the presence of an extensive communication device $\Lambda$, providing advices $\lambda$ according to some distribution $p(\lambda)$, the possible payoffs $U_i$ can then be written as
\begin{eqnarray}
\label{eq: payoff decomposition}
U_i = \sum_{a,\theta,\lambda}p(\lambda)p(\theta)\prod_{j=1}^n \sigma_j(a_j|a_{j-1,\dots,j-m_j},\lambda) u_i(a,\theta),
\end{eqnarray}
on which we use the fact that here we are considering the advice as being independent of the types of the players \cite{Bernhard2008},  that is, $p(\lambda,\theta)=p(\lambda)p(\theta)$. In the context of Bell's theorem, this is equivalent to the measurement independence (or ``free-will'' assumption) \cite{Chaves2015,hall2010local,barrett2011much}.

Since $\lambda$ is a classical variable, we can consider that, given $\lambda$, the behaviours $\sigma_i$ are deterministic functions of the available history $h_{i}$ and the type $\theta_i$. Thus
\begin{equation}
\sigma_i(a_i|a_{i-1,\dots,i-m_i,\lambda})=
\delta_{a_i,f_{i,\lambda}(a_{i-1,\dots,i-m_i},\theta_i)}. 
\end{equation}
This implies that the set of payoffs vectors $(U_1,\dots,U_n)$ form a convex polytope $\mathcal{U}_{E}$, being characterized by sets of linear inequalities given by
\begin{eqnarray}
\label{eq:payoffineq}
\sum_{j=1}^n\beta_j U_j\leq \beta_0.
\end{eqnarray}

To make the connection of multistage games with a Bell scenario we first introduce the standard Bell scenario, consisting of $n$ distant parties. Each party $i$ upon receiving their parts of a joint physical system make measurements indexed by inputs $\theta_i$ obtaining outcomes $a_i$. The Bell experiment is then described by a joint probability distribution $p(a\vert \theta)=p(a_1,\dots,a_n \vert \theta_1,\dots,\theta_n)$ of  measurement outcomes conditioned on the kind of measurements being performed.  In a classical description, the source of correlations between the distant parties can be described by a random variable $\Lambda$ governed by a distribution $p(\lambda)$. Since the parties are distant (space-like separated) and further assuming the independence between the inputs and source of correlations $\Lambda$, one obtains the paradigmatic local hidden variable model decomposition given by
\begin{equation}
\label{eq:LHV}
p(a\vert \theta) = \sum_{\lambda}p(\lambda) p(a_1 \vert \theta_1, \lambda)\dots p(a_n \vert \theta_n, \lambda).
\end{equation}

\begin{figure}
    \centering
    \includegraphics[scale = 0.15]{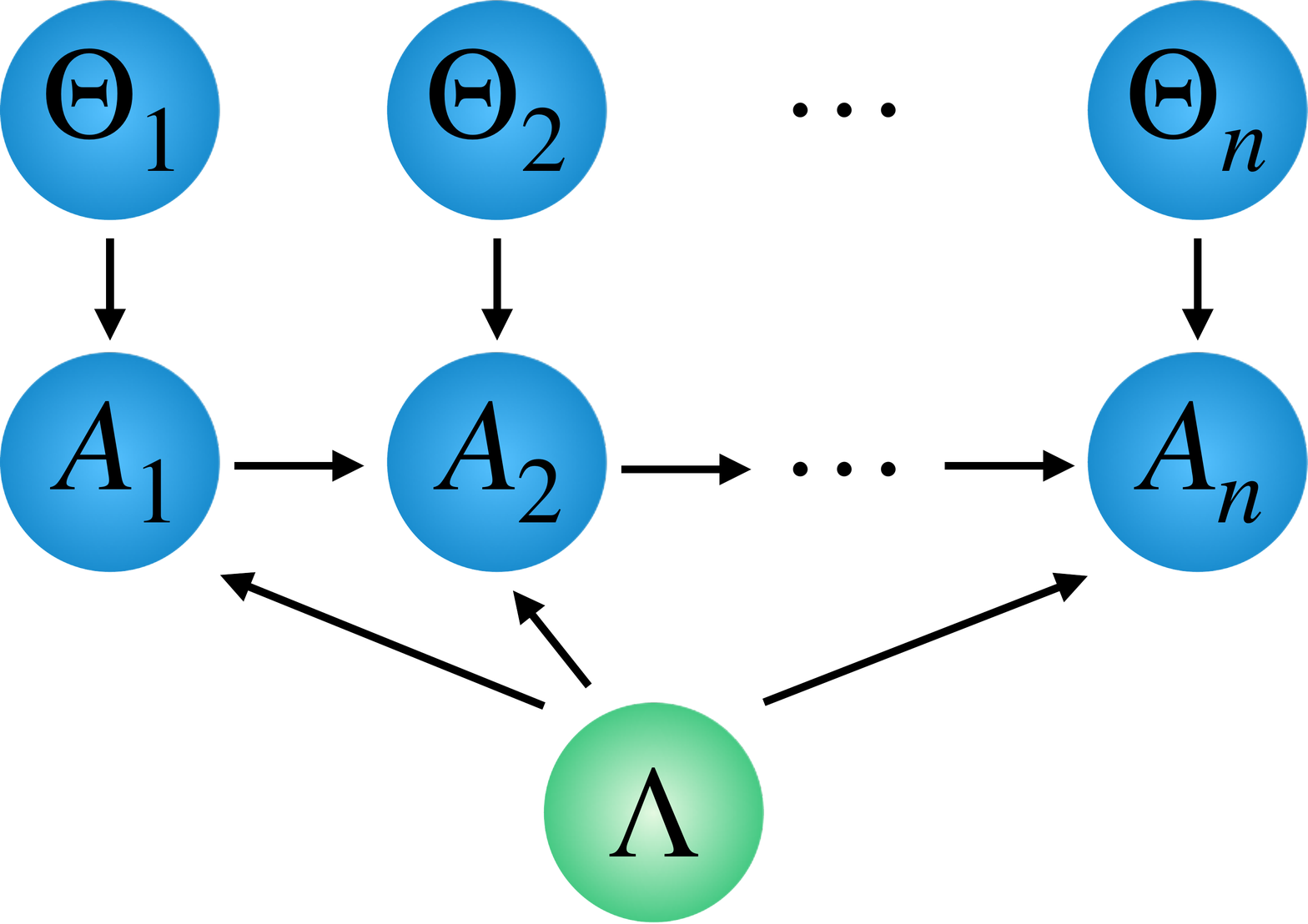}
    \caption{Multistage game with a limited memory $m_i=1$ (thus with $h_{i}={a_{i-1}}$) represented here as a directed acyclic graph.}
    \label{fig: Causal m=1}
\end{figure}

More generally, one can consider variations of a Bell scenario where communication between the parties, either of the inputs or outputs, is considered \cite{bacon2003bell,pironio2003violations,pawlowski2010non,maxwell2014bell,Chaves2015,ringbauer2016experimental,Brask2017}. In particular, similarly to the multistage game, we can consider a sequential scenario where party $i$ may receive previous measurement outcomes. Classically, considering that party $i$ receives the previous $m_i$ measurement outcomes, the conditional probability distribution can then be written as 
\begin{equation}
\label{eq:SLHV}
p(a|\theta) =\sum_{\lambda}p(\lambda)\prod_{i=1}^n p_i(a_i|a_{i-1,\dots,i-m_i},\theta_i,\lambda) .
\end{equation}
Similarly to the case of multistage games, in a classical description the probability of obtaining a certain measurement outcome can be written in terms of deterministic functions such that $p_i(a_i|a_{i-1,\dots,i-m_i},\theta_i)=\delta_{a_i,g_{i,\lambda}(a_{i-1},\dots,a_{i-m},\theta_i)}$. Furthermore, just as the set of payoff functions \eqref{eq: payoff decomposition}, the hidden variable decomposition in \eqref{eq:SLHV} forms a polytope as well. It can be characterized by a set of linear inequalities, known as Bell inequalities and generally written as
\begin{equation}
\label{eq: Bell inequality}
\sum_{a,\theta}\alpha_{a,\theta}p(a|\theta) \leq L,
\end{equation}
where $L$ is the bound respected by any classical distribution of the form \eqref{eq:SLHV}. A graphical way of representing a given Bell scenario is via a directed acyclic graph \cite{pearl2009causality}. In this graph each node represents a variable of interest (inputs, actions or hidden variables) and the directed arrows account for the cause and effect relations among them. For instance, Fig. \ref{fig: Causal m=1} shows an example associated with a multistage game with $n$ players and limited memory ($m_i=1$), in which case each player has only access to the actions of the player immediately before.

A formal link between multistage games and a Bell scenario with communication of outputs can then be made by setting
\begin{eqnarray}
\label{eq: key association}
\alpha_{a,\theta} = p(\theta)u_i(a,\theta).
\end{eqnarray}
That is, we are identifying the coefficients of a Bell inequality with the prior distribution and the payoff function of player $i$ in the multistage game. Substituting this into the Bell inequality we obtain
\begin{equation}
\label{eq: inequality}
\sum_{a,\theta}p(\theta)p(a|\theta)u_i(a,\theta)= U_i \leq L.
\end{equation}
That is,  when considering a classical advisor, the average payoff $U_i$ of player $i$ is limited by the classical bound $L$ defining the Bell inequality for the scenario with outcome communication. Similarly, the space of payoff vectors defined by \eqref{eq:payoffineq} will also be limited by this classical bound.

As a particular case of this result we can retrieve the connection made in Ref. \cite{Brunner2013} between Bayesian games and the standard Bell scenario without communication of measurement outputs (setting $m_i=0$). In this case, the game features a canonical correlation device instead of an extensive one. Thus the possible payoff functions are of the form
\begin{eqnarray}
U^{C}_i = \sum_{\lambda}\sum_{a,\theta}p(\lambda)p(\theta)\prod_{i=1}^n \delta_{a_i,f_{i,\lambda}(\theta_i,\lambda)}u_i(a,\theta),
\end{eqnarray}
which trivially implies that the space of payoff vectors obtained via a canonical communication device $\mathcal{U}_C$ is contained in  the space $\mathcal{U}_E$ obtained using an extensive communication device. In fact, given the correspondence between games with incomplete information and Bell scenarios, it is not difficult to prove that, for certain games, $\mathcal{U}_C$ is a proper subset of $\mathcal{U}_E$, that is, $\mathcal{U}_C \subset\mathcal{U}_E$. For that aim it is enough to find one example of a Bell inequality valid for a canonical communication device and that can be violated by an extensive communication device. Consider for instance the paradigmatic CHSH inequality \cite{CHSH}, described in terms of the probability $p(a_1,a_2\vert \theta_1,\theta_2)$ as
\begin{eqnarray}
& & p(0,0\vert 0,0) +p(0,0\vert 0,1)+p(0,0\vert 1,0)\\ \nonumber
& & -p(0,0\vert 1,1) -p_{A_1}(0\vert 0) -p_{A_2}(0\vert 0) \leq 0
\end{eqnarray}
where $p_{A_1}(0\vert 0)=p(0,0\vert 0,0)+p(0,1\vert 0,0)$ and similarly for $A_2$. If the payoff function is described by the CHSH inequality via the rule \eqref{eq: key association}, this means that using a CCD this payoff function will be always smaller than zero (corresponding to the bound of the CHSH inequality). However, using an ECD, the action/outcome of the first player is accounted for in the second player's decision and in this case is able to violate the CHSH inequality up to its maximum and achieve a payoff given by $1/2$.

In the following, we will consider the quantum realization of a multistage game and the new forms of equilibrium it entails.

\section{Quantum equilibrium}
\label{sec:4}
If the advisor is allowed to provide quantum advice we can define quantum canonical communication devices (QCCD) as well as quantum extensive communication devices (QECD). In order to be aligned with the classical realizations of an advice where the advisor sets the profile to be played, we assume that a quantum advice consists on both a quantum system shared by the parts and the optimal POVMs to be performed by them. These assumptions are important to decide whether a point is an equilibrium point but do not alter in any way the quantum payoff space, which is the main focus of this work.

The QCCD and QECD will generate spaces of possible payoffs $\mathcal{U}^{(Q)}_C$ and $\mathcal{U}^{(Q)}_E$, respectively. Similarly to the result proven above for classical strategies, we will also prove that $\mathcal{U}^{(Q)}_C \subset\mathcal{U}^{(Q)}_E$ 
even though in some particular games the space of payoff functions will be the same.

Moreover, as proven in Ref. \cite{Brunner2013}, it follows that $\mathcal{U}^{(Q)}_C \supset \mathcal{U}_C$. Here we will prove a similar result for multistage games, that is, $\mathcal{U}^{(Q)}_E \supset \mathcal{U}_E$. Surprisingly, however, is the fact that $\mathcal{U}^{(Q)}_C\not\subset\mathcal{U}_E$. For those aims, given the link between Bell inequalities and the payoff functions established in \eqref{eq: key association} and \eqref{eq: inequality}, it will suffice to consider Bell inequalities which can be violated by quantum correlations.

We will mostly consider symmetric games where all players have the same payoff function [all of them associated to a Bell inequality via \eqref{eq: key association}]. Nonetheless, we also provide an example showing that our results also hold for the asymmetric case.
For symmetric games and under the assumption that the advisor provides the quantum state and the measurements leading to the optimal violation of the corresponding Bell inequality with that state, any distribution $p(a|x)$ violating the Bell inequality defining a given game will also constitute a Nash equilibrium point. That follows because given the quantum state shared by the advisor any other choice of measurements by the players will not increase their payoffs. Furthermore, when dealing with symmetric games, we will focus on the maximum quantum violation (represented by the probability distribution $p^*(a|x)$) of a given Bell inequality as it provides the optimal payoff function of the game. It reveals not only a new equilibrium, as would any other nonlocal point, but also the most desirable outcome. If one is interested in the physical implementation of the game, there will always be the presence of noise, meaning that the maximal theoretical violation will not be achieved. Nevertheless, as long as the corresponding Bell inequality is violated, not only we will have a new Nash equilibrium but also a better payoff as compared to the case of a classical advise.

The optimal quantum distribution is obtained by a collection of local measurements performed by each of the players leading to
\begin{eqnarray}
\label{eq:quantumpstar}
p^*(a,\theta) = \tr\left[\left(\bigotimes_{i=1}^n M_{a_i}^{(x_i)}\right)\rho\right],
\end{eqnarray}
where $\rho$ is the quantum state shared between the players by the advisor and the measurements are parametrized by the inputs $x_i$ and corresponding outcomes $a_i$ and thus represented by POVM operators $M_{a_i}^{(x_i)}$ respecting  $\sum_{a_i}M_{a_i}^{(x_i)}=\Id$. To make the connection with the multistage game we associate the measurements inputs $x_i$ with the pair $(\theta_i,a_{i-1,\dots,i-m_i})$ and the actions according to the measurement outcomes of the game $a_i$. 

As a first side remark, we notice that one can also extend the notions above to the case where the parties share post-quantum correlations fulfilling the non-signalling conditions \cite{popescu1994quantum}. For simplicity, consider the bipartite case where the parties share correlations described by the probability distribution $q(a_1,a_2 \vert x_1,x_2)$ respecting the non-signalling constraints $q(a_1\vert x_1)= \sum_{a_2}q(a_1,a_2 \vert x_1,x_2)=\sum_{a_2}q(a_1,a_2 \vert x_1,x^{\prime}_2)$ and $q(a_2\vert x_2)= \sum_{a_1}q(a_1,a_2 \vert x_1,x_2)=\sum_{a_1}q(a_1,a_2 \vert x^{\prime}_1,x_2)$. One possible analogous of equation \eqref{eq:quantumpstar} would be given by
\begin{equation}
p(a_1,a_2\vert x_1,x_2)=q(a_1,a_2 \vert x_1,f(x_2,a_1)),    
\end{equation}
where $f$ is a function mapping $(x_2,a_1)$ to a variable of the same cardinality as $x_2$.
We also highlight the fact that if one is dealing with payoff functions corresponding to full-correlator Bell inequalities (as it will be the case in the examples we provide below), then it is sufficient to consider only the non-signalling set of correlations, because such correlations already achieve the maximum violation of full-correlator Bell inequalities.

As a second side remark, we notice that in order to find the optimal classical as well as quantum payoffs (associated the classical and quantum bounds of the corresponding Bell inequality) one could employ the graph-theoretical approach proposed in Refs. \cite{cabello2014graph,acin2015combinatorial} and recently generalized in Ref. \cite{poderini2020exclusivity} to the case where the actions/outcomes of a given player might depend on the outcomes/actions of previous players. In this graph-theoretical framework the classical and quantum bounds correspond to properties of a graph and could be used, for instance, to decide whether a given payoff function might have a quantum advantage or not. As exemplified below, that approach will not be necessary for our purposes, but remains as an interesting venue for future investigations.

\subsection{QCCD ties with QECD that beats ECD}
\label{sec: QCCD ties with QECD that beats ECD}

We start considering a multistage game consisting of two players only. The simplest scenario for which we could find a quantum strategy beating the classical one is obtained when each of the player have three possible types/inputs and binary actions/outcomes, that is,  $\Theta_1 = \Theta_2 = \{0,1,2\}$, $A_1 = A_2 = \{0,1\}$, and $p(\theta_1) = p(\theta_2) = \frac{1}{3}$. It can be seen as a particular case of the causal structure in Fig. \ref{fig: Causal m=1}.

Given the connection \eqref{eq: inequality}, the payoff provided by classical advisor in the CCD or ECD form will always be bounded by the classical bound of the associated Bell inequality, that in this case can be generally written as
\begin{eqnarray}
\label{eq:general2}
\sum_{a_1,a_2,\theta_1,\theta_2}\alpha_{a_1,a_2,\theta_1,\theta_2}p(a_1,a_2|\theta_1,\theta_2)\leq L.
\end{eqnarray}
Of particular relevance is the Bell inequality (bounding the classical correlations of the scenario considered here and illustrated in Fig. \ref{fig: Causal m=1}) introduced in Ref. \cite{Chaves2015}, given by
\begin{multline}
\label{eq:ineqchav}
\langle\,(A_0 - A_2)\otimes (B_0 - B_2)\,\rangle \\ - \langle\, (A_1 - A_2) \otimes (B_1 - B_2)\, \rangle \leq 4,
\end{multline}
where $\langle A_i \otimes B_j \rangle \coloneqq \sum_{a_1,a_2} (-1)^{a_1 + a_2} p(a_1,a_2\vert \theta_i,\theta_j)$. This inequality can be readily written in the form \eqref{eq:general2}, corresponding, via the association \eqref{eq: key association}, to a payoff function given by
\begin{equation}
\label{eq: payoff two parts symmetric}
u_i(a,\theta) = -9[(\theta_1 \oplus_3 \theta_2) -1 ](-1)^{a_1 + a_2},
\end{equation}
in which $\oplus_3$ represents sum modulo $3$.

Given the connection \eqref{eq: inequality}, the maximum payoff provided by a classical advisor, either in CCD or ECD forms, will be $U_i=4$. Quantum mechanically, however, this inequality can be violated up to a maximum of $3\sqrt{3} \approx 5.1962$. Interestingly this maximum value is the same for both QCCD and QECD advisors, and thus corresponds to a new game equilibrium in both these cases.

This value can be analytically obtained with the following quantum strategy. Assume that the advisor distributes a singlet state between the participants, $|\psi^-\rangle = (|01\rangle - |10\rangle)/\sqrt{2}$. Assume now rank-1 projective measurements for both parties, such that $M^{(\theta_i)}_{a_i=0} - M^{(\theta_i)}_{a_i=1} = \boldsymbol{v}^{(i)}_{\theta_i} \cdot \boldsymbol{\sigma},\quad (i=1,2)$ (that is, we are explicitly considering a QCCD strategy, as the previous outcomes are not used by the second player.). For each player, take now the vectors $\boldsymbol{v}^{(i)}_{\theta_i}$ as vertices of an equilateral triangle, in the form
\begin{align}
\boldsymbol{v}^{(1)}_{0} &= \boldsymbol{k}, \\
\boldsymbol{v}^{(1)}_{1} &= \frac{-\boldsymbol{k} + \sqrt{3}\,\boldsymbol{k}_\perp}{2}, \\
\boldsymbol{v}^{(1)}_{2} &= \frac{-\boldsymbol{k} - \sqrt{3}\,\boldsymbol{k}_\perp}{2},\\
\boldsymbol{v}^{(2)}_{0} &= \frac{-\boldsymbol{k}_\perp - \sqrt{3}\boldsymbol{k}}{2}, \\
\boldsymbol{v}^{(2)}_{1} &= \boldsymbol{k}_\perp, \\
\boldsymbol{v}^{(2)}_{2} &= \frac{-\boldsymbol{k}_\perp + \sqrt{3}\boldsymbol{k}}{2},
\end{align}
with $\boldsymbol{k}, \boldsymbol{k}_\perp$ being any unit vectors such that $\boldsymbol{k}\cdot\boldsymbol{k}_\perp = 0$. Direct computation shows that the violation $3\sqrt{3}$ is attained with this choice.

To prove that the value achieved by this specific strategy is indeed optimal for the modified and traditional Bell scenarios, one can cast the associated optimization problem as a semi-definite program (SDP). The brute-force  optimization of quantum operators $M^{(x_i)}_{a_i}$ is a computationally hard task, but upper bounds for the optimal payoffs with quantum advisors can be found efficiently with the use of an SDP. This can be done, for instance, with the NPA hierarchy method \cite{NPA1,NPA2} and its adaptation to scenarios with communication \cite{Moreno2020}.  Adaptation to scenarios with communication is done by considering the pertinent scenario as a particular case of an augmented nonsignaling scenario. Thus, for example, a quantum realization for $P(a_1,a_2\vert \theta_1,\theta_2)$ as $\Tr[M^{\theta_1}_{a_1} \otimes M^{\theta_2,a_1}_{a_2}\,\rho]$ would only be considered possible if positive semidefinite operators $M^{x}_{a},\,M^{y,y^\prime}_{b},\,\rho$, with $\sum_a M^{x}_a = \mathbbm{1}$, $\sum_b M^{y,y^\prime}_b = \mathbbm{1}$, and $\Tr[\rho]=1$, can be found, allowing the nonsignaling distribution $P_{0}(a,b\vert x,y,y^\prime) = \Tr[M^x_a \otimes M^{y,y^\prime}_b \rho]$, such that the distribution of interest is obtained as the particular case $y^\prime = a_1$, i.e., $P(a_1,a_2\vert \theta_1,\theta_2) = P_0(a_1,a_2\vert \theta_1,\theta_2,a_1)$.

Whenever the upper bound found by the SDP coincides with the analytical bound given by a specific shared quantum state and quantum measurements, we have then a proof of its optimality. That is precisely the case for the scenario considered here, for which the SDP returns the same value $3\sqrt{3}$, regardless of using a QCCD or QECD.

If the parties can share non-signalling (post-quantum) correlations, then one can achieve the algebraic maximum violation of inequality \eqref{eq:ineqchav} to the algebraic maximum payoff given by $
6$.

\subsection{QCCD beats ECD}

As the next scenario, we consider a general multistage game between $n$ players, each having two possible types/inputs and actions/outcomes, that is, $\#A_i = \#\Theta_i = 2$ for all $i$. Without loss of generality we are going to label those as $A_i=\{0,1\}$ and $\Theta_i = \{0,1\}$. Furthermore we will consider general cases of games that can be mapped to a \textit{partially paired graph}  \cite{Jones2005}. That is, there are at least two players $i$ and $j$ such that no player in the game, including themselves, access information of their actions simultaneously. 

\begin{figure}
    \centering
    \includegraphics[scale=0.15]{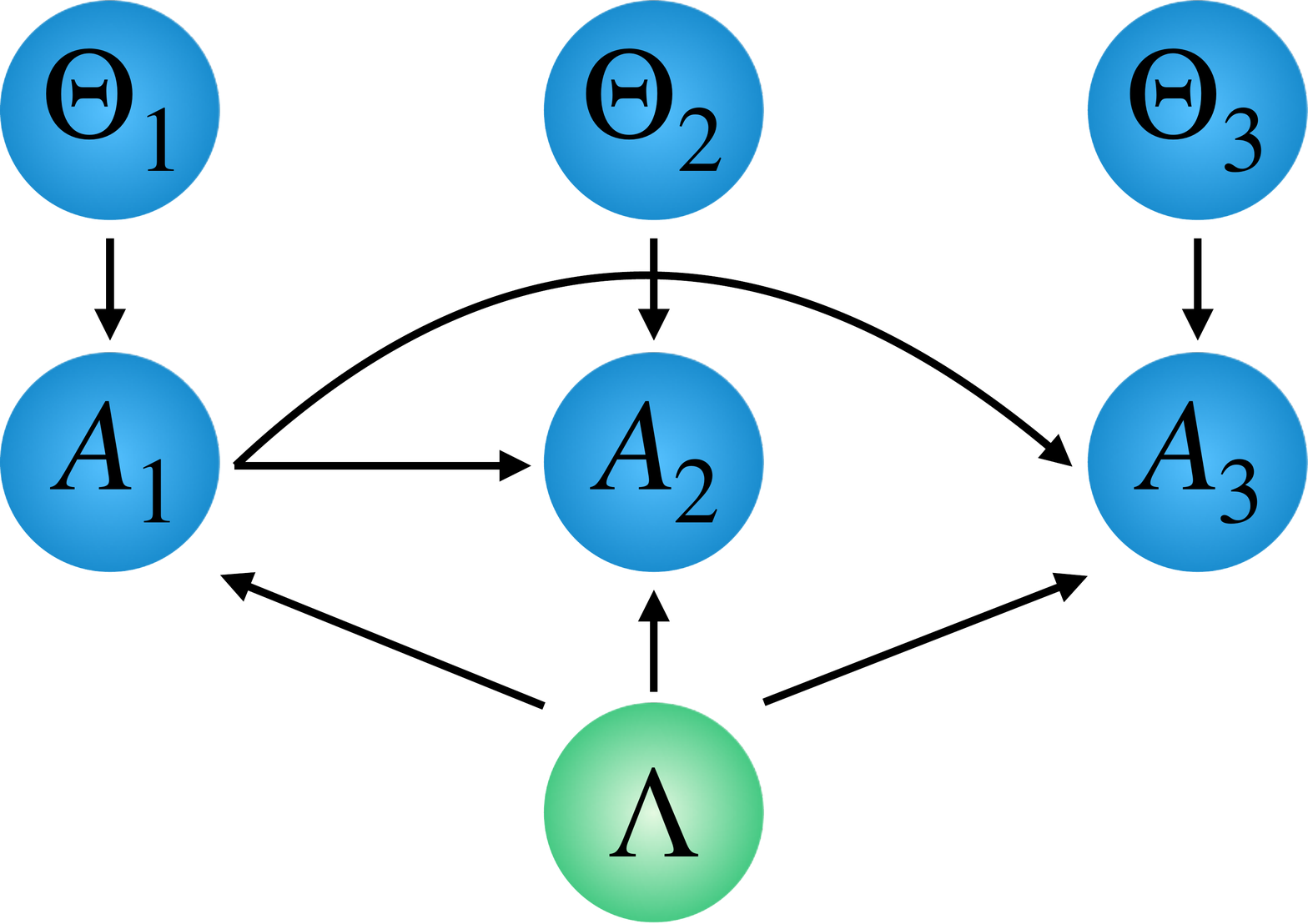}
    \caption{Bell scenario associated with a tripartite multistage game in which the last player cannot access the action of his immediate predecessor.}
    \label{fig: tripartite}
\end{figure}

We have seen a particular case of that in Fig. \ref{fig: Causal m=1}, corresponding to $m_i=1$. Another possibility in the tripartite scenario is the one depicted in Fig. \ref{fig: tripartite}, where players 2 and 3 get the action/outcome $a_1$ but player 3 has no information about $a_2$.

All games that can be mapped to a partially paired graph have been proven to be bounded (classically) by the n-partite Svetlichny inequality given by \cite{Seevinck2002}
\begin{eqnarray}
\label{eq: Svetlichny}
\nonumber
S_n = \sum_{a, \theta}(-1)^{\left(\sum_j^n \theta_j\right)\left(1 + \sum_j^n \theta_j\right)/2 + \sum_{l=1}^n a_l}p(a|\theta)\leq 2^{n-1}.\\
\end{eqnarray}
Via the relation \eqref{eq: key association} this sets the payoff functions of the players to be given by
\begin{eqnarray}
\label{eq: Svetlichny's payoff}
u_i(a,\theta) = (-1)^{\left(\sum_j^n \theta_j\right)\left(1 + \sum_j^n \theta_j\right)/2+\sum_{l=1}^n a_l}\left(p(\theta)\right)^{-1}.
\end{eqnarray}
Thus any classical strategy is limited to a maximum payoff $U_i =2^{n-1}$.

A QCCD strategy, can nonetheless surpass the maximum classical value and establish a new equilibrium. The quantum maximum achieved by a QCCD strategy is given by $U_i=2^{n-1}\sqrt{2}$ \cite{Seevinck2002}. For that, the players share an $n$-partite GHZ state
\begin{eqnarray}
|GHZ\rangle = \frac{1}{2}\left(|0\rangle^{\otimes n}+|1\rangle^{\otimes n}\right),
\label{eq:GHZ_npartite}
\end{eqnarray}
and perform the association between behaviours and measurements as
\begin{eqnarray}
\nonumber
\sigma_i(a_i|\theta_i) = \tr\left[M_{a_i}^{\theta_i}|GHZ\rangle\langle GHZ|\right],
\end{eqnarray}
in which
\begin{eqnarray}
M_0^{\theta_1} - M_1^{\theta_1} = \frac{1}{\sqrt{2}}\left(\sigma_y+(-1)^{\theta_1}\sigma_x\right)
\label{eq:meas_svet_MA}
\end{eqnarray}
and, for $i\neq1$,
\begin{eqnarray}
M_0^{\theta_i} - M_1^{\theta_i} = - \theta_i\sigma_y - (1 \oplus \theta_i)\sigma_x
\label{eq:meas_svet_MB}
\end{eqnarray}
$\sigma_x$ and $\sigma_y$ representing the Pauli matrices.

This choice of behaviours constitute an equilibrium outcome which beats the classical one, returning a larger payoff. It thus forbids any rational player in a game with a QCCD to deviate from it, once it represents the maximal violation of the Svetlichny inequality achievable by a quantum nonsignaling distribution.

If the parties can share non-signalling (post-quantum) correlations, then one can achieve the algebraic maximum violation of the Svetlichny inequality leading to a payoff of $2^n$.

\subsection{QECD beats QCCD}

To show the existence of scenarios in which QECD beats QCCD we consider the particular case of the previous section in which $n=3$ and  $m_1=m_2=m_3=1$, that is, the participants only access the action/outcome of the previous player (see Fig. \ref{fig: Causal m=1}).

We start by considering the QCCD case. By following the optimal strategy described in equations (\ref{eq:GHZ_npartite})-(\ref{eq:meas_svet_MB}) with $n=3$ parties, we achieve the maximum Svetlichny inequality violation by a nonsignaling scenario, given by $S_3=4\sqrt{2}$, also corresponding to the optimal value for the payoff. In the particular case of three parties, the fact that this value correspond to the optimum has been proven in \cite{Mitchell2004}, but can also be easily proven by the semi-definite approach described above.

Turning our attention to the QECD case, it follows that higher payoffs can be attained. The SDP approach puts an upper bound of $S_3=6$ to the Svetlichny's inequality violation. By brute-force numerical optimization, we found a tripartite qubit-state and measurements achieving this maximum. The quantum state and the measurements are given by
\begin{align}
\rho &= \frac{1}{2}\left( |\Psi_1\rangle\langle \Psi_1| + |\Psi_2 \rangle\langle \Psi_2 | \right), \\
(M_A)^{\theta_1=0}_{a_1=1} &= (M_A)^{\theta_1=1}_{a_1=0} = \lambda\,\left(\frac{\mathbbm{1} + \boldsymbol{a}\cdot \boldsymbol{\sigma}}{2}\right),\\
(M_B)^{\theta_2,\, a_1}_{a_2} &= \frac{\delta_{a_2, \neg\, a_1} + \delta_{a_2, \theta_2}}{2}\,\mathbbm{1} - (-1)^{a_2}\,\frac{ \theta_2 \oplus a_1\oplus 1}{2}\, \boldsymbol{b}\cdot \boldsymbol{\sigma}, \\
(M_C)^{\theta_3, a_2}_{a_3} &= \begin{cases}
\displaystyle (a_3\oplus 1)\,\mathbbm{1},\quad \textrm{if } \theta_3 = a_2 \\
\\
\displaystyle \frac{\mathbbm{1} + (-1)^{a_3 + a_2}\boldsymbol{c}\cdot\boldsymbol{\sigma}}{2},\quad \textrm{otherwise} \\
\end{cases}
\end{align}
where $\lambda \approx 0.4989$, and $\boldsymbol{a},\,\boldsymbol{b},\,\boldsymbol{c}$ are unit vectors given approximately by $[-0.4368,\,-0.3031,\,-0.8469]^T$,  $[-0.2855,\,-0.9581,\,0.0214]^T$, $[0.1391,\,0.9556,\,-0.2599]^T$, respectively, where $T$ indicates transposition. In turn, the states that compose $\rho$ are given by
\begin{equation}
|\Psi_1\rangle\! =\! \begin{bmatrix}
\phs 0.0609 - 0.0704\,i \\
-0.0296 + 0.6751\,i \\
-0.0421 - 0.6386\,i \\
\phs 0.1811 - 0.1257\,i \\
\phs 0.0260 - 0.0067\,i \\
-0.1178 + 0.1548\,i \\
\phs 0.0949 - 0.1579\,i \\
\phs 0.0634 + 0.0000\,i
\end{bmatrix}\!\!,
\end{equation}
and
\begin{equation}
|\Psi_2\rangle\! =\! \begin{bmatrix}
\phs 0.6644 - 0.0953\,i \\
\phs 0.0685 - 0.0543\,i \\
\phs 0.0211 + 0.2330\,i \\
\phs 0.5264 - 0.3653\,i \\
\phs 0.1728 + 0.0865\,i \\
\phs 0.0251 - 0.0016\,i \\
-0.0332 + 0.0586\,i \\
\phs 0.1845 + 0.0000\,i\end{bmatrix}.
\end{equation}

Interestingly the shared state achieving the maximum payoff for a QECD strategy is a mixed state while the measurements are non-projective.

In this case, if the advice is mediated by (post-quantum) correlations, then one can achieve the algebraic maximum violation of the tripartite Svetlichny inequality leading to a payoff of $8$.

\subsection{Example of an asymmetric game}

In order to provide an example of an asymmetric multistage game, here we follow a similar idea to the one employed in \cite{Brunner2013}. An asymmetric game is characterized by different players presenting different payoff functions.

Let us build our example in a scenario similar to that of section \ref{sec: QCCD ties with QECD that beats ECD}: two players for which $\Theta_1 = \Theta_2 = \{0,1,2\}$, $A_1 = A_2 = \{0,1\}$, and $p(\theta_1) = p(\theta_2) = \frac{1}{3}$. The difference here being the payoff functions. For player $1$ we set:
\begin{eqnarray}
\nonumber
u_1(a,\theta) = -(-1)^{a_1 + a_2} 9(1 - \theta_1)^2(1 - \theta_2)^2[(\theta_1\oplus_3\theta_2) - 1],
\end{eqnarray}
and for player $2$:
\begin{eqnarray}
\nonumber
u_2(a,\theta) = -(-1)^{a_1 + a_2} 36 \left(\frac{1}{2}\right)^{\theta_1+\theta_2}\theta_1\theta_2[(\theta_1\oplus_3\theta_2) - 1].
\end{eqnarray}

The mean payoff of each player reads:
\begin{eqnarray}
\left\{\begin{array}{cc}
     U_1 = & \langle A_0\otimes B_0\rangle - \langle A_0 \otimes B_2\rangle - \langle A_2\otimes B_0\rangle \\
     U_2 = & \langle A_1\otimes B_2\rangle + \langle A_2 \otimes B_1\rangle - \langle A_1\otimes B_1\rangle
\end{array}\right. . 
\end{eqnarray}

While $U_1$ and $U_2$ are always smaller than $3$, the sum of the payoffs $U_1 + U_2$ is limited by $4$ when the advice is classical. This clearly shows a conflict of interests between the participants: for a player to raise its payoff to the maximum value possible, the other participant receives only a third of his allowed payoff. A quantum advice as described in section \ref{sec: QCCD ties with QECD that beats ECD} can raise this value up to $3\sqrt{3}$, which reveals an enlargement of the space of payoff functions that takes place when using an QECD or an QCCD instead of an ECD. As a side remark, if we go further and consider a postquantum advisor \cite{popescu1994quantum} the sum of the payoffs can reach the maximum value of $6$, thus optimising the individual payoffs up to its maximum possible value, implying a Nash equilibrium which has the interesting feature to align the interests of the participants, otherwise in conflict.

Finally, we also mention briefly how to obtain other paradigmatic kind of game, known as a zero-sum game. For that aim all we have to do is set $u_2(a,\theta) = -u_1(a,\theta)$, and
\begin{eqnarray}
u_2(a,\theta) = -9[(\theta_1 \oplus_3 \theta_2) -1 ](-1)^{a_1 + a_2},
\end{eqnarray}
which is the same as in equation \eqref{eq: payoff two parts symmetric}. Once more, a classical advise limits the payoffs to $U_i\leq4$, while a quantum advisor provides the possibility of larger payoffs, $U_i\leq3\sqrt{3}$. Again, putting in evidence the enlargement of the payoff space as a consequence of replacing an ECD by a QECD. 

\section{Discussion}
\label{sec:5}
In this work we have established a formal link between multistage games with incomplete information and Bell scenarios with outcome communication. Every multistage game is faithfully represented by the same causal structure representing hidden variable models where the measurement outcome of a given player might be communicated for other players in its future light cone. Thus, the payoff functions are restricted by such classical hidden variable models. In particular, every Bell inequality characterizing a given multistage game/Bell scenario can be associated with a payoff function in such a way that a Bell inequality violation obtained with measurement on entangled states not only lead to new  payoffs but sometimes to new correlated equilibrium in the game.

Our results can be seen as a generalization of the connection between Bell's theorem and Bayesian games made in Ref. \cite{Brunner2013} to an extended scenario, where the parties are not imposed to be space-like separated and thus can communicate its measurement outcomes. As such, our results are achieved in a device-independent setting, not relying on any specific constraints such as in the early works on quantum game theory \cite{Eisert1999}. Thus it provides an interesting application and new venue of research for causal structure beyond the paradigmatic case of local hidden variable models. A particularly interesting venue would be to consider the relation between Bell scenario relaxing the measurement independence (``free-will'') assumption and the kind of games discussed by Forges \cite{Forges2006} where the advice might be correlated with the types of the players. Finally, machine learning has been recently shown to provide an alternative manner to characterize Bell correlations as well as optimize Bell inequalities \cite{PhysRevLett.120.240402,PhysRevLett.122.200401,krivachy2020neural,bharti2019teach,doi:10.1116/5.0007529}. Given the connection between Bell scenarios and game theory, another relevant path for future research is to understand whether machine learning might also be applied to the analysis of equilibrium points and optimal payoffs.

\section{Acknowledgements}
We thank an anonymous Referee for a detailed report leading us to include the example of an asymmetric game as well as improving the presentation of the paper.
We acknowledge the John Templeton Foundation via the Grant Q-CAUSAL No. 61084, the Serrapilheira Institute (Grant No. Serra-1708-15763), the Brazilian National Council for Scientific and Technological Development (CNPq) via the National Institute for Science and Technology on Quantum Information (INCT-IQ) and Grants No. 307172/2017-1 and No. 406574/2018-9, the Brazilian agencies MCTIC and MEC. 

\bibliographystyle{apsrev4-2}

\end{document}